\documentclass[journal]{IEEEtran}
\usepackage{cite}
\usepackage{times}
\usepackage{amsmath}
\usepackage{amssymb}
\usepackage{epsfig}
\usepackage{caption}
\usepackage{graphicx}
\usepackage{epstopdf}
\usepackage{subfigure}
\usepackage{multirow}
\usepackage{siunitx}
\usepackage{bm}
\usepackage{csquotes}
\usepackage{braket}
\usepackage{color}

\allowdisplaybreaks
\newcommand{\figref}[1]{{Fig.~\ref{#1}}}

\usepackage{soul}

\ifCLASSINFOpdf

\else

\fi

\begin{document}
%
\title{Valley-Hall topological plasmons in a graphene nanohole plasmonic crystal waveguide}

\author
{
Jian Wei~You,~\IEEEmembership{Member,~IEEE,}
Zhihao~Lan,~
Qiaoliang~Bao,~
and~Nicolae C.~Panoiu,~\IEEEmembership{Member,~IEEE}
\thanks {Manuscript received xxx, 2020. This work was supported by the European Research Council (ERC), Grant Agreement no. ERC-2014-CoG-648328. (Corresponding author: Nicolae C. Panoiu.)}
\thanks{J. W. You, Z. Lan, and N. C. Panoiu are with the Department of Electronic and Electrical Engineering, University College London, London, WC1E7JE, United Kingdom. (e-mail: j.you@ucl.ac.uk, z.lan@ucl.ac.uk, and n.panoiu@ucl.ac.uk)}
\thanks{Q. Bao is with the Department of Materials Science and Engineering, and Australian Research Council Centre of Excellence in Future Low-Energy Electronics Technologies, Monash University, Clayton, Victoria 3800, Australia. (e-mail: qiaoliang.bao@monash.edu)}
\thanks{Color versions of one or more of the figures in this paper are available online
at http://ieeexplore.ieee.org.}
\thanks{Digital Object Identifier 10.1109/JSTQE.2020.XXXX.}
}

\markboth{IEEE Journal of Selected Topics in Quantum Electronics,~Vol.~XX, No.~YY, ZZ~2020}%
{Shell \MakeLowercase{\textit{et al.}}: Valley-Hall topological plasmons in graphene nanohole
plasmonic crystal waveguide}

\maketitle

\begin{abstract}
We demonstrate that unidirectional and backscattering immune propagation of terahertz optical waves can be
achieved in a topological valley-Hall waveguide made of graphene nanohole plasmonic crystals. In
order to gain deeper physical insights into these phenomena, the band diagram of graphene nanohole
plamsonic crystals has been investigated and optimized. We found that a graphene plasmonic crystal
with nanohole arrays belonging to the $C_{6v}$ symmetry group possesses gapless Dirac cones, which
can be gapped out by introducing extra nanoholes such that the symmetry point group of the system
is reduced from $C_{6v}$ to $C_{3v}$. Taking advantage of this feature, we design a mirror
symmetric domain-wall interface by placing together two optimized graphene plasmonic crystals so as
to construct valley-polarized topological interface modes inside the opened bandgap. Our
computational analysis shows that the valley-Hall topological domain-wall interface modes can be
achieved at an extremely deep subwavelength scale, and do not rely on the application of external
static magnetic fields. This work may pave a new way to develop highly-integrated and robust
terahertz plasmonic waveguides at deep-subwavelength scale.
\end{abstract}


\IEEEpeerreviewmaketitle

\section{Introduction}

\IEEEPARstart{T}{opological} photonics has recently attracted a great deal of attention, chiefly
due to the unique opportunities it provides to manipulate light in a robust way, immune to
structural imperfections and backscattering \cite{opag19RMP}. In \cite{hr08PRL}, an analogue of the
quantum Hall edge states in photonic crystals containing time-reversal-breaking magneto-optical
media was proposed theoretically for the first time. Soon afterwards, such topological edge modes
were observed directly in microwave experiments \cite{wcjs09Nature}. Since then, the photonic
topological states have been extensively studied both from a fundamental science viewpoint
\cite{ljs14NatPhot,ljs16NatPhy,ks17NatPhot,lgyz17PRL}, as well as from the perspective of potential
applications
\cite{wcjm08PRL,ypwg13APL,mffm14PRL,hmfm13NatPhot,cjnm16NatMat,bnve17Sci,sggl17NatPhot}. For
instance, a number of topological photonic waveguides \cite{wcjm08PRL,ypwg13APL,mffm14PRL} have
been developed to reduce the back-reflection in the presence of inherent structural disorder. In
addition, some other topological photonic devices, such as robust delay lines
\cite{hmfm13NatPhot,cjnm16NatMat} and topological lasers \cite{bnve17Sci,sggl17NatPhot}, have also
been proposed and demonstrated experimentally.

One of the prerequisites of the development of efficient on-chip photonic integrated circuits at
the nanoscale, is achieving a tight optical field confinement beyond the diffraction limit of
light. Graphene \cite{ngmj04Science,ngmj05Nature,gpn12NatPhot} has been shown to be a promising
platform to facilitate reaching this goal, due to the fact that its physical properties can be
easily controlled by voltage-tuning or chemical doping. Moreover, graphene plasmons, the energy
carriers in graphene based photonic systems, exhibit a tight optical field confinement and have
long intrinsic relaxation times in the terahertz and up to mid-infrared frequency range
\cite{trvi14JSTQE,avfr14JSTQE,wlgp15NatMat,ylhc18NatPhot,nmzw18Nature}. Indeed, it has been shown
that a periodically patterned graphene nanostructure under a static magnetic field can host
topological one-way edge plasmons up to infrared frequencies
\cite{jcsf17PRL,pyxa17NatComm,ylp20sa}, owing to the breaking of the time reversal symmetry.

Topological transport of plasmons in graphene nanostructures can also be realized in a time
reversal symmetric manner by exploiting the idea of valley-contrast transport, via electrically
biased patterned metagates \cite{jfs18PRL} or inversion symmetry broken honeycomb lattices of
periodically arranged graphene nanodisks \cite{qlqc17OE,qqrw19JPDAP}. Recently, a large area of
periodic graphene nanohole arrays with a period down to \SI{100}{\nano\meter} has been fabricated
\cite{sintou10JACS,ymlz14Nanoscale,gpmr18NanoLett}. Compared to the crystals of periodically
arranged graphene nanodisks, graphene nanohole crystals offer the particularly useful feature of
being an electrically connected surface, yet the band structure of the graphene plasmons can be
easily engineered in a purely geometric manner \cite{ycys14NanoLett}. To the best of our knowledge,
topological plasmons in graphene nanohole crystals have not been achieved in the absence of a
static magnetic field.

In this work, we propose a novel valley-Hall topological graphene plasmonic crystal waveguide
comprised of periodically patterned air nanoholes. To achieve this goal, we firstly construct an
inversion-symmetry protected Dirac cone by creating a $C_{6v}$-symmetric array of nanoholes into a
graphene sheet. Subsequently, in order to break the spatial inversion symmetry, we introduce
additional nanoholes with different radius into this graphene nanohole crystal. In this way, the
gapless Dirac cone is gapped out and consequently a topological bandgap emerges. In order to study
the topological features of this novel graphene nanohole crystal containing two kinds of nanoholes,
the radius of the additional set of nanoholes is optimized so as to maximize the width of the
emerging topological bandgap. Taking advantage of this wide topological bandgap, we place two
optimized graphene nanohole crystals together in a mirror symmetric manner, so as to construct a
domain-wall interface, which gives rise to a valley-Hall plasmonic edge mode inside the topological
bandgap.

Employing comprehensive numerical simulations, the valley-Hall plasmon propagation along this
carefully-designed interface has been studied, and some typical topological propagation features
have been observed, including unidirectional backscattering-immune propagation. Moreover, we have
explored the influence of the intrinsic loss of graphene on the characteristic loss propagation
length of valley-Hall topological plasmons in this graphene nanohole crystal waveguide, too, and
found out that the plasmon lifetime is a key parameter in determining the plasmon propagation
performance. Compared to previously studied topological magneto-plasmons, the proposed valley-Hall
topological plasmons investigated in this work are time-reversal-invariant and do not depend on
external static magnetic fields. Therefore, our proposal would greatly simplify experimental
implementations of topological graphene plasmons and could be extremely useful for the development
of robustly integrated on-chip photonic devices.

\section{The aim of the study and its key ideas}\label{sec:PhyComp}

\begin{figure}[t!]
\centering\includegraphics[width=\columnwidth]{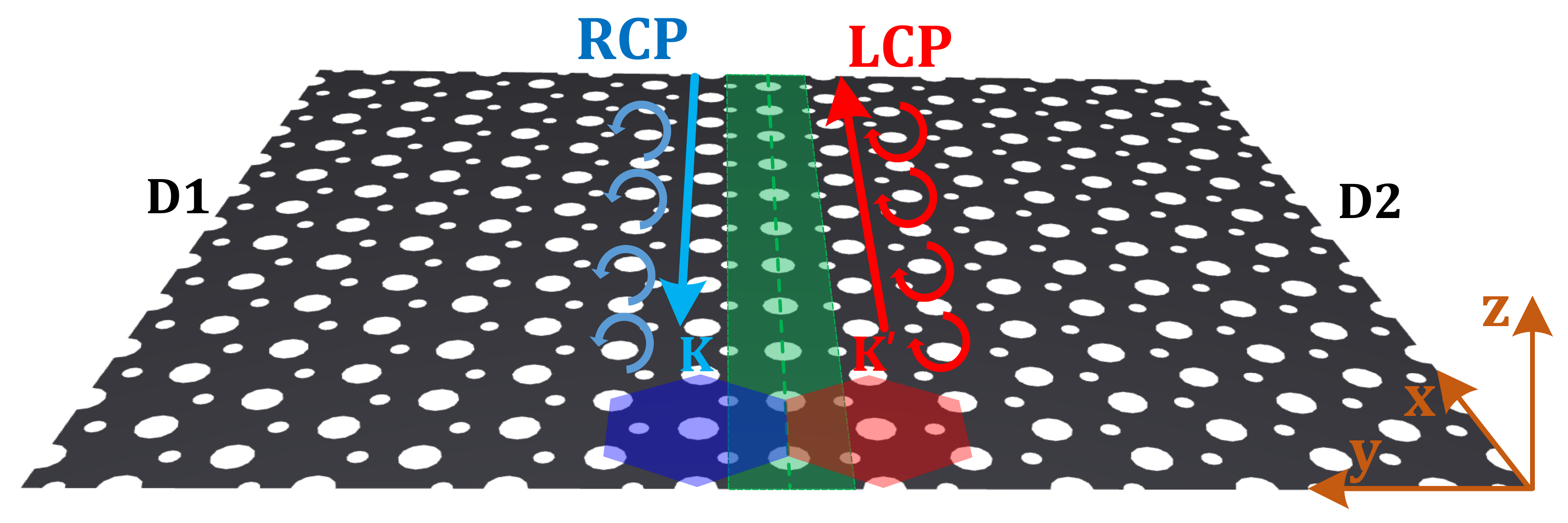} \caption{ Schematic of a topological
valley-Hall waveguide supporting valley-contrast unidirectional transport of terahertz plasmonic
waves, i.e., RCP modes at the $K$ valley points and LCP modes at the $K'$ valley points. The
waveguide consists of a domain-wall interface constructed by placing together two mirror-symmetric
graphene nanohole plasmonic crystals (domains $D1$ and $D2$), whose unit cells are highlighted by
blue and red hexagons, respectively.} \label{fig:TopoValleySchematic}
\end{figure}

The aim of this study is to design a topological waveguide for terahertz plasmonic waves based on
quantum valley-Hall like effects using graphene nanohole plasmonic crystals (see the illustration
in \figref{fig:TopoValleySchematic}). The waveguide consists of a domain-wall interface, which is
highlighted in \figref{fig:TopoValleySchematic} by a transparent green strip along the $x$-axis,
and it is constructed by joining together two properly-designed graphene nanohole crystals referred
to as domains $D1$ and $D2$. The $D1$ domain is the mirror symmetric counterpart of the $D2$
domain, with respect to a plane passing through the domain-wall interface and perpendicular onto
the $y$-axis. The unit cells of the $D1$ and $D2$ domains are marked by blue and red hexagons,
respectively. Each hexagonal unit cell consists of three small nanoholes and three large nanoholes
etched in a graphene sheet and the unit cells are arranged periodically in a hexagonal lattice. The
array of large nanoholes has $C_{6v}$ symmetry and in the absence of the small nanoholes the
structure has spatial-inversion-symmetry protected Dirac cones. In order to gap out these Dirac
cones, additional small nanoholes are introduced in the unit cell to break the spatial inversion
symmetry and reduce the system symmetry from $C_{6v}$ down to $C_{3v}$.

Since the unit cell of the $D1$ domain is a mirror symmetric counterpart of that of the $D2$ domain
with respect to the domain-wall interface, the band diagram of the $D1$ crystal is identical to
that of the $D2$ crystal. This key feature is important in our design, as it is particularly useful
that the $D1$ and $D2$ crystals have a common bandgap. Moreover, since in the real space the $D1$
domain is a mirror symmetric counterpart of the $D2$ domain, in momentum space the first Brillouin
zone of the $D1$ domain can be obtained from that of the $D2$ domain by a rotation of $180$
degrees, namely the $K$-valley points of the $D1$ domain correspond to the $K'$-valley points of
the $D2$ domain. As we will see later on, the valley mode at $K$ points is right-hand circularly
polarized (RCP), whereas the valley mode at $K'$ points is left-hand circularly polarized (LCP).
This indicates that the valley states are inherently chiral, i.e., the chirality of the $K$-valley
is opposite to that of the $K'$-valley. Due to this important feature, a chirality-momentum locked
edge mode will emerge in the topological bandgap. In the following, we will discuss and analyze
these ideas in more detail through full-model calculations of the band diagrams and propagation
properties of the topological interfacial waveguide mode.

\section{Results and Discussion}\label{sec:ResDis}
In this section we present the main results of our study. In particular, we first discuss the band
structure of the graphene plasmonic crystals used in our work, as well as the mechanism through
which an interfacial topological plasmonic crystal waveguide can be formed and its specific
waveguiding properties.

\subsection{Band diagrams of graphene nanohole crystals}\label{sec:BandInfi}
\begin{figure}[t!]\centering
\includegraphics[width=\columnwidth]{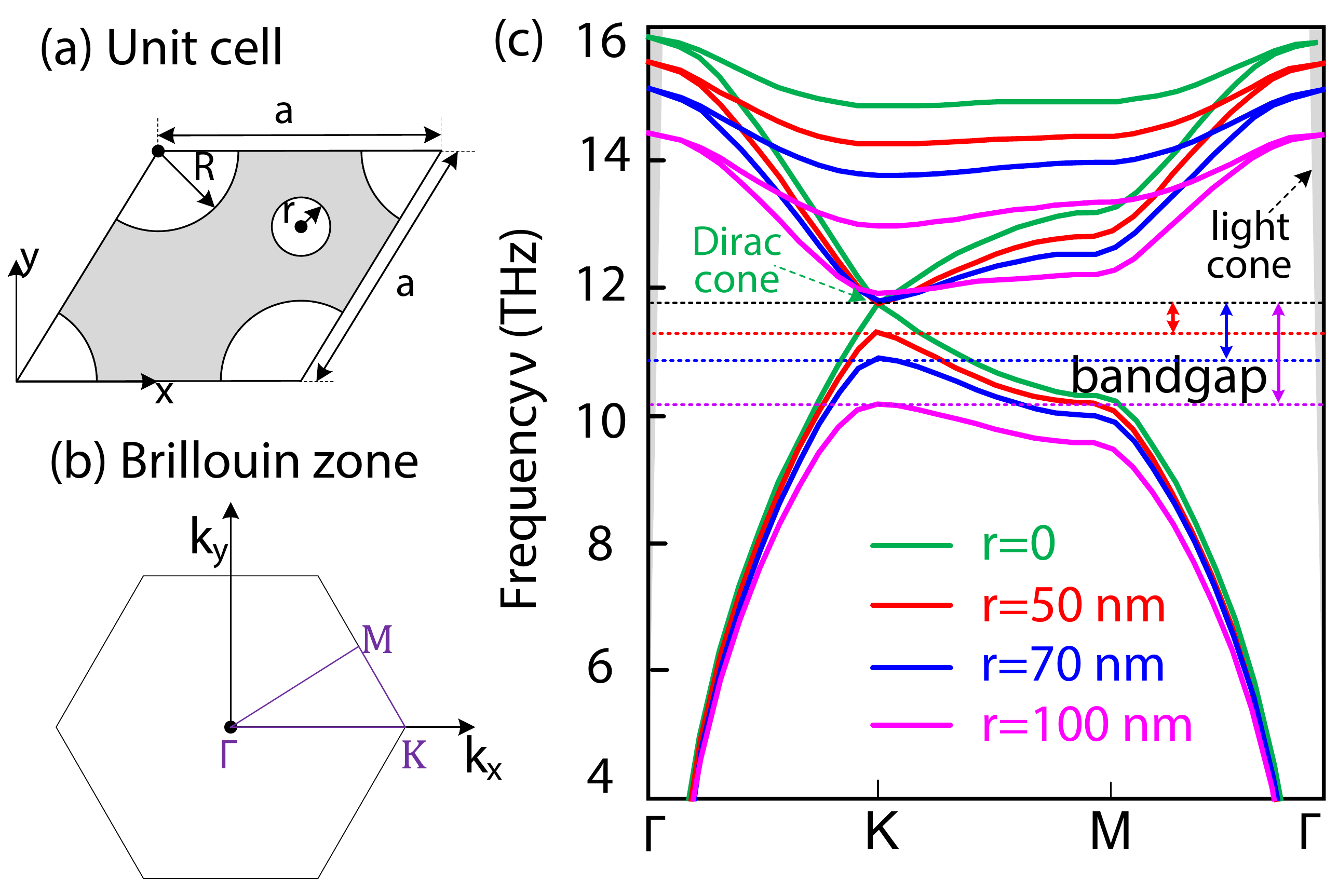}
\caption{Band diagrams of graphene nanohole plasmonic crystals. (a) Primitive unit cell consisting
of two nanoholes with radius $R$ and $r$, etched in a graphene sheet (grey shaded region). (b)
First Brillouin zone (FBZ) of the graphene nanohole plasmonic crystal, where the high symmetry
points have been labeled as $\Gamma$, $K$, and $M$. (c) First three bands of the graphene nanohole
plasmonic crystal, calculated for different values of the radius of the smaller holes, $r$. In
these calculations, $a = \SI[parse-numbers = false]{400\sqrt{3}}{\nano\meter}$ and
$R=\SI{140}{\nano\meter}$. The numerical calculations show that the width of the bandgap increases
with the radius $r$. More importantly, the bandgaps exist at a deep subwavelength scale, around
\SI{11.5}{\tera\hertz} (the operating wavelength is about \num{40} times larger than the lattice
constant).} \label{fig:BandDiagInfCell}
\end{figure}

As shown in \figref{fig:BandDiagInfCell}(a), the unit cell of the proposed graphene plasmonic
crystal contains two nanoholes with differen size. Moreover, the holes are arranged periodically in
a hexagonal lattice, whose first Brillouin zone (FBZ) is depicted in
\figref{fig:BandDiagInfCell}(b). In this study, the lattice constant and the radius of the large
holes are fixed as $a = \SI[parse-numbers = false]{400\sqrt{3}}{\nano\meter}$ and
$R=\SI{140}{\nano\meter}$, respectively.

The optical properties of graphene are primarily characterized by its surface conductivity, which
is generally given by the Kubo's formula \cite{gp16book,ybbp19Nanophoton}. Within the random-phase
approximation \cite{Hans08jap,wzg15tnano}, this formula can be expressed as the sum of inter-band
and intra-band contributions, $\sigma_s = \sigma_{intra}(\omega,\mu_c,\tau,T) + {\sigma
_{inter}}({\omega,\mu_c,\tau,T})$. To be more specific, the intra-band part is given by:
\begin{equation}\label{eq:sintra}
\sigma_{intra} = \frac{e^2 k_B T\tau}{\pi\hbar^2\left(1-i\omega\tau\right)}\left[ \frac{\mu _c}{k_B
T} + 2\ln \left(e^{-\frac{\mu_c}{k_B T}} + 1\right) \right],
\end{equation}
where $\mu _c$ is the chemical potential, $\tau$ is the relaxation time, $T$ is the temperature,
$e$ is the electron charge, $k_B$ is the Boltzmann constant, and $\hbar$ is the reduced Planck's
constant. Moreover, as long as $\mu _c \gg k_B T$, which usually holds at room temperature, the inter-band
part can be approximated as:
\begin{equation}\label{eq:sinter}
\sigma_{inter} = \frac{ie^2}{4\pi\hbar}\ln \left[ \frac{2\left|\mu_c\right| - (\omega +
i\tau^{-1})\hbar}{2\left|\mu_c\right| + (\omega + i\tau^{-1})\hbar} \right].
\end{equation}

In this study, unless otherwise specified, we use $\mu_c=\SI{0.2}{\electronvolt}$,
$T=\SI{300}{\kelvin}$, and $\tau=\SI{50}{\pico\second}$
\cite{la14AcsNano,dyml10NatNano,yllz12NanoLett,pwbp13PRL}, whereas the influence of the relaxation
time on the propagation characteristics of the topological mode will be discussed at the end of the
paper. Here, we should note that the band diagram of graphene plasmonic crystal is not dependent on the relaxation time of graphene. The only influence of this physical parameter is that the width of each band increase as the relaxation time increases, as proved in \cite{jcsf17PRL}.

We have employed the finite-element method (FEM) in Comsol Multiphysics 5.4 \cite{ComsolSoft} to
compute the band diagrams of graphene nanohole plasmonic crystals with different values of the
radius of the small nanoholes, and the corresponding results are summarized in
\figref{fig:BandDiagInfCell}(c). As we have explained above, when $r=0$, the large holes form a
periodic array with $C_{6v}$ symmetry, which possesses inversion-symmetry-protected Dirac cones
located at the $K$ point. As shown in \figref{fig:BandDiagInfCell}(c), the first band and the
second band linearly cross each other at the $K$ point, at the frequency of \SI{11.8}{\tera\hertz}.
Different from most traditional all-dielectric photonic crystals
\cite{smns17NatPhot,hlyq19NatComm}, the gapless Dirac cone of this graphene nanohole plasmonic
crystal exists at a frequency far below the air light cone, which is marked by the grey shaded
region in \figref{fig:BandDiagInfCell}(c). As a consequence, the corresponding topological plasmons
can be formed at extremely deep subwavelength scale. In our case, the plasmon wavelength is around
$40$ times larger than the lattice constant.

In order to open up the inversion-symmetry-protected Dirac cone, small nanoholes are added to the
plasmonic crystal, which means in this case $r\neq0$. In this way, the symmetry point group of the
unit cell shown in \figref{fig:BandDiagInfCell}(a) is reduced from $C_{6v}$ to $C_{3v}$.
Consequently, the $C_{6v}$-symmetry-protected Dirac cone is gapped out, owing to the breaking of
the spatial inversion symmetry. Moreover, the numerical results presented in
\figref{fig:BandDiagInfCell}(c) show that the width of the bandgap increases when the radius of the
small nanoholes increases. To be more specific, the frequency bandwidth of the first bandgap
increases from \SI{1}{\tera\hertz} to about \SI{2}{\tera\hertz} when the radius of the small
nanoholes increases from \SIrange{50}{100}{\nano\meter}.
\begin{figure}[t!]\centering
\includegraphics[width=\columnwidth]{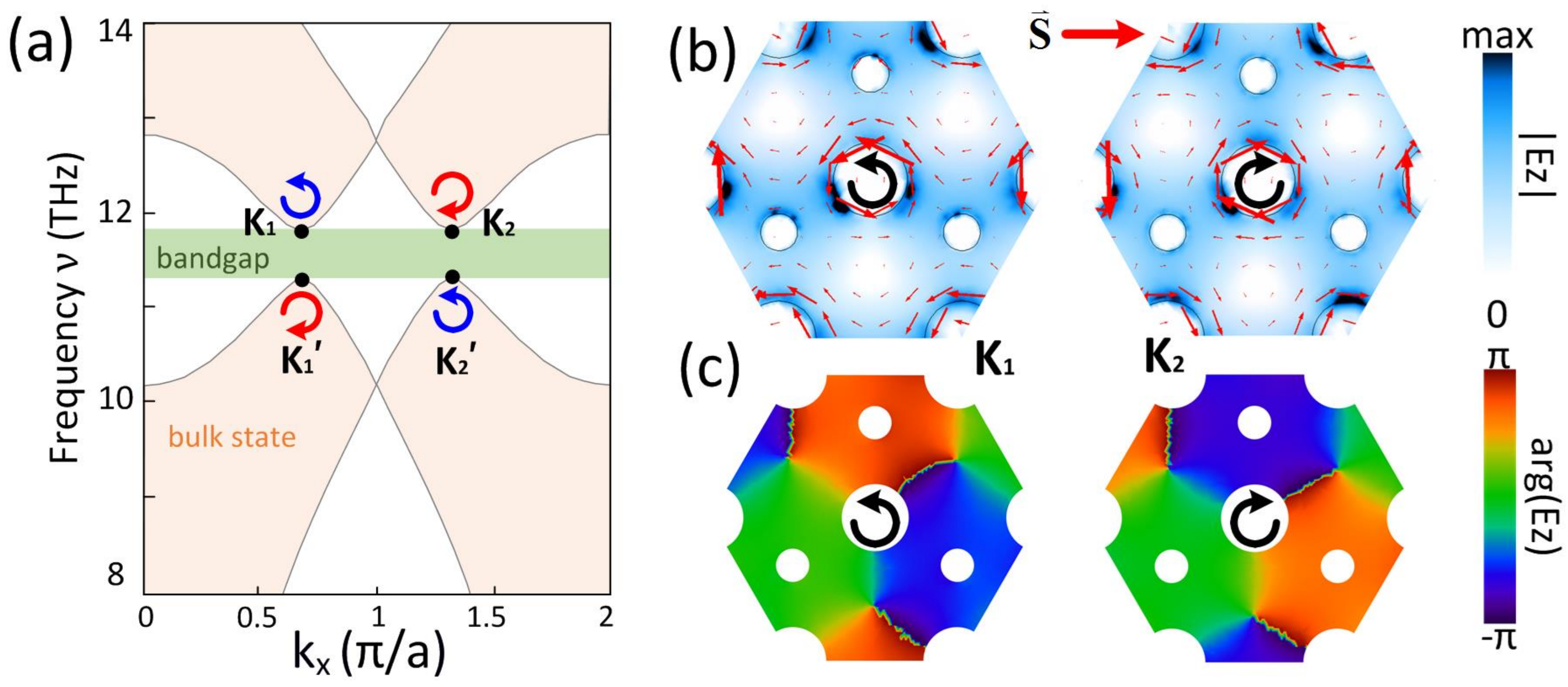}
\caption{Chirality properties of the bulk valley modes of graphene nanohole plasmonic crystals. (a)
Projection of the band diagram of the graphene nanohole plasmonic crystal along the $\Gamma$-$K$
direction, where the four bulk valley modes at $K_1$, $K_2$, $K_1'$ and $K_2'$ are marked by dots.
The chirality of the four valley modes are marked by clockwise- or counterclockwise-oriented
arrows. (b) Electric field and Poynting vector distributions in a unit cell, determined in the
$x-y$ plane for the $K_1$ and $K_2$ valley modes. A tight field confinement is suggested by the
optical field distribution, whereas the distribution of the Poynting vector (red arrows) shows
either a RCP or a LCP energy flow. (c) Phase distributions of the $E_z$ component in a unit cell,
computed in the $x-y$ plane for the $K_1$ and $K_2$ valley modes, which show that the two valley
modes have opposite chirality, i.e., for the $K_1$ and $K_2$ valley modes the $E_z$ phase vortex
rotates either anticlockwise or clockwise around the center of the unit cell, respectively.}
\label{fig:ValleyChirality}
\end{figure}

The bulk valley modes obtained by gapping out the Dirac cone have nontrivial topological
properties, which are responsible for the emergence of valley-Hall like topological domain-wall
interface modes to be discussed in the next subsection. To illustrate this phenomenon, we present
in \figref{fig:ValleyChirality}(a) the projected band diagram of the graphene nanohole plasmonic
crystal. In this figure, the beige shaded regions represent the projection of the bulk modes, and
the bandgap is marked by a green shaded region. The four bulk valley modes are labeled $K_1$,
$K_2$, $K_1'$, and $K_2'$, where each valley mode has an intrinsic vorticity with the corresponding
chirality marked by either a clockwise- or counterclockwise-oriented arrow. In practical
applications, these optical vortices can be used to selectively excite different valley modes by
choosing the specific angular momentum of the excitation source. To be more specific, if we input
light with the frequency equal to the frequency of the upper valley modes, only RCP light could
couple efficiently to the valley mode at $K_1$, whereas only LCP light couples to the valley mode
at $K_2$.

In order to demonstrate this valley-chirality-locking property, the electric field and Poynting
vector distributions of $K_1$ and $K_2$ valley modes are depicted in
\figref{fig:ValleyChirality}(b), where the colormap denotes the amplitude of $E_z$ component and
the Poynting vector is depicted by red arrows (note that the $K_1'$ and $K_2'$ valley modes show
similar behaviors, thus not shown). It can be seen in this figure that the electric field is
tightly confined around the large nanoholes, thus illustrating the plasmon-induced near-field
confinement effect. Moreover, the distribution of the Poynting vector shows that the chirality of
the valley modes at $K_1$ and $K_2$ are RCP and LCP, respectively. Furthermore, this
valley-chirality-locking feature also manifests through the corresponding distribution of the phase
of $E_z$, as depicted in \figref{fig:ValleyChirality}(c). To be more specific, the phase
distribution of $E_z$ component at the $K_1$ valley winds counterclockwise from $\pi$ to $-\pi$
with respect to the center of the unit cell. On the contrary, the phase of $E_z$ at the $K_2$
valley point winds clockwise. Relying on this valley-chirality locking property, a valley-Hall
topological graphene nanohole plasmonic crystal waveguide will be designed and studied in the
following subsections.

\subsection{Band diagrams of graphene plasmonic crystal waveguides}\label{sec:BandFini}
\begin{figure}[t!]\centering
\includegraphics[width=\columnwidth]{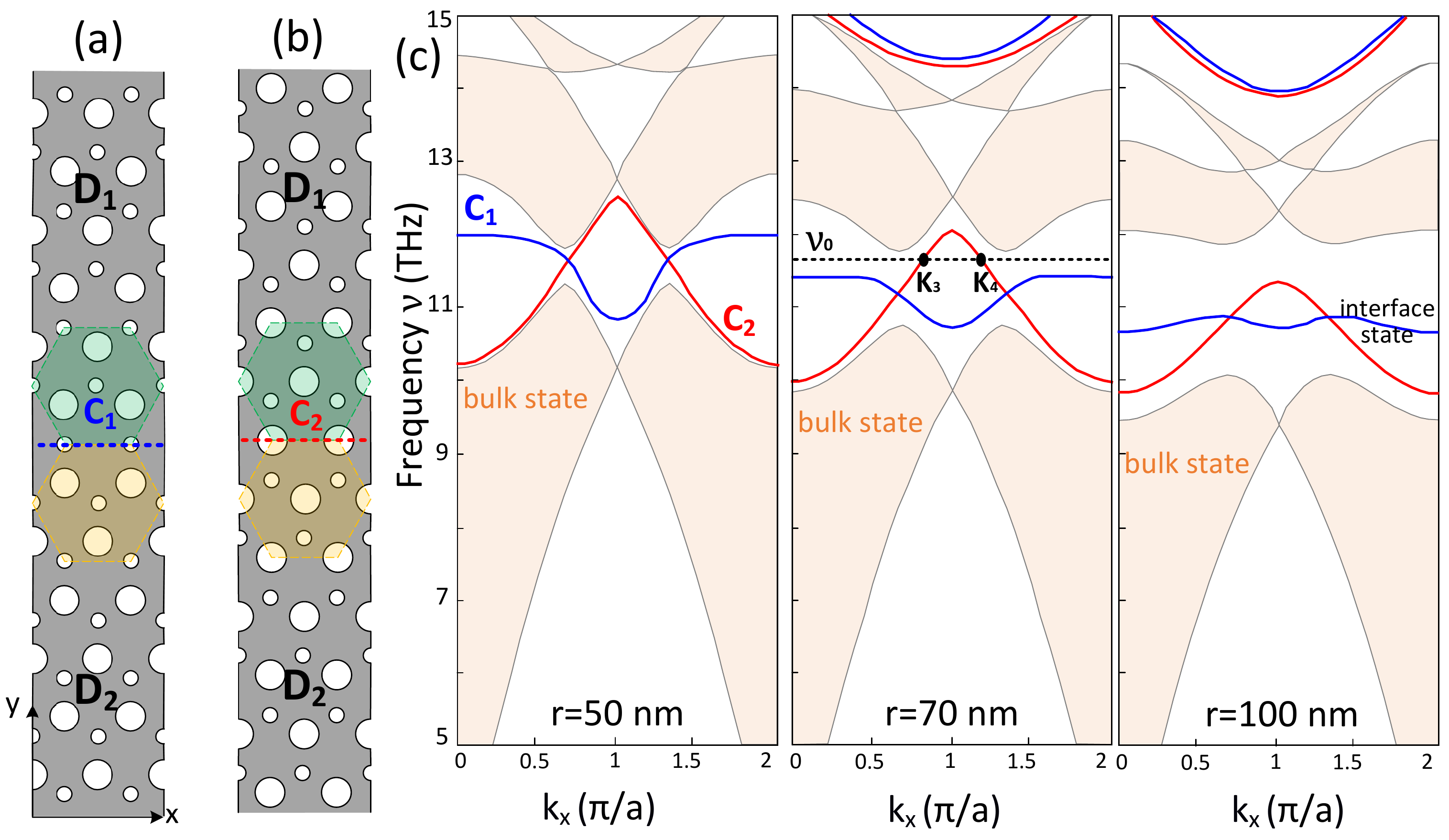}
\caption{Topological edge states of graphene nanohole plasmonic crystal waveguides. (a), (b) Two
complementary domain-wall interfaces $C_1$ (blue dashed line) and $C_2$ (red dashed line),
respectively, which are constructed by placing together two mirror-symmetric domains, $D1$ and
$D2$. (c) Band diagrams of graphene nanohole plasmonic crystal waveguides, determined for
$r=$~\SIlist{50;70;100}{\nano\meter}. Due to the presence of the domain-wall interfaces, two
edge-mode bands are formed inside the topological bandgap. The blue (red) solid line corresponds to
the edge-mode band of the domain-wall interface $C_1$ ($C_2$).} \label{fig:BandDiagInterface}
\end{figure}

So far, we have explained how one can design a graphene nanohole plasmonic crystal that possesses a
relatively wide bandgap. In order to construct valley-Hall topological edge modes inside this
bandgap, we place together two optimized graphene nanohole plasmonic crystals to form a topological
domain-wall interface, as illustrated in \figref{fig:BandDiagInterface}(a) and
\figref{fig:BandDiagInterface}(b). In our study, the $D2$ domain is a mirror symmetric counterpart
of the $D1$ domain. There are two key reasons why one constructs a valley-Hall topological
domain-wall interface in this way.

First, apart from a symmetry transformation, the unit cell of the $D1$ domain is identical to that
of the $D2$ domain. Therefore, one can avoid bandgap mismatch issues, namely the bandgaps of the
plasmonic crystals in the domains $D1$ and $D2$ around the interface are perfectly matched. Second,
in order to achieve a chirality-momentum locking feature of the edge modes supported by the
domain-wall interface, the chirality of the $D1$ domain should be opposite to that of the $D2$
domain. This can be achieved by simply rotating in the momentum space by $180$ degrees the unit
cell of the $D2$ domain or, equivalently, by ensuring that in the real space the domains $D1$ and
$D2$ can be obtained from each other through a mirror symmetry transformation with respect to the
interface. In other words, the $K$-valley points of the $D1$ domain overlap with the $K'$-valley
points of the $D2$ domain. Owing to this unique feature, the chirality of the modes in the
nontrivial bandgap of the $D1$ domain is opposite to that of the modes in the (same) nontrivial
bandgap of the $D2$ domain.

Since the domain-wall interface is constructed by placing together the $D1$ and $D2$ domains, it
possesses a pair of counter-propagating edge modes with opposite chirality. Here, we should note
that the domain-wall interface $C_1$ in \figref{fig:BandDiagInterface}(a) is a complementary
counterpart of the domain-wall interface $C_2$ presented in \figref{fig:BandDiagInterface}(b),
namely the two configurations can be obtained from each other by interchanging the positions of the
large and small nanoholes. The structures depicted in \figref{fig:BandDiagInterface}(a) and
\figref{fig:BandDiagInterface}(b), are periodic along the $x$-axis but are finite along the
$y$-axis and symmetric with respect to the interface. Therefore, the topological features of the
domain-wall interface $C_1$ are complementary with respect to those of the domain-wall interface
$C_2$. To simplify the discussion, in the following we only focus on the topological properties of
the domain-wall interface $C_2$.

Using the FEM solver in Comsol Multiphysics 5.4, the band diagrams of the graphene nanohole
plasmonic crystal waveguides depicted in \figref{fig:BandDiagInterface}(a) and
\figref{fig:BandDiagInterface}(b) were calculated, and the corresponding results are given in
\figref{fig:BandDiagInterface}(c). Due to the presence of the domain-wall interfaces, two
additional valley-Hall topological edge-mode bands are formed inside the first bandgap of the
graphene nanohole plasmonic crystal. Different from the topological edge-mode bands induced by the
breaking of the time-reversal symmetry, these valley-Hall topological edge-state bands induced by
the breaking of the spatial-inversion symmetry do not connect the conduction and valence bulk bands
of the plasmonic crystal. In \figref{fig:BandDiagInterface}(c), the blue (red) solid line
corresponds to the edge-state band of the domain-wall interface $C_1$ ($C_2$), whereas the beige
shaded regions indicate the projection of the bulk modes. Furthermore, these band diagrams also
suggest that the absolute value of the group velocity of the edge-state bands inside the
topological bandgap decreases when the radius $r$ increases.

As it is well-known, as a consequence of slow-light effects, the smaller the absolute value of the
group velocity is, the less confined the optical field will be at the domain-wall interface.
Therefore, in order to avoid slow-light effects and the narrow operational bandwidth associated
with flatten bands, the radius $r$ should be as small as possible. However, in most practical
applications, it is desirable to use nanoholes with radius $r$ as large as possible, as this would
make the fabrication processes less challenging and would also lead to a wider bandgap. After a
careful consideration of both these factors, we found out that the optimum value of the radius $r$
was \SI{70}{\nano\meter}. Therefore, in the following analysis we used this value for $r$.

More specifically, when $r=\SI{70}{\nano\meter}$, the edge-state mode (red solid line) has a
relatively large group velocity and is located in a topological bandgap with a rather large width
of \SI{2}{\tera\hertz}. As illustrated in \figref{fig:BandDiagInterface}(c), if one excites this
properly designed graphene nanohole plasmonic crystal waveguide using a monochromatic light source
with frequency $\nu_0$, one can generate two distinct waveguide modes corresponding to the $K_3$
and $K_4$ points. More importantly, the group velocity of the edge-mode band at the $K_3$ point is
positive, whereas the corresponding group velocity at the $K_4$ point is negative. As the
propagation direction of a waveguide mode is determined by the sign of the corresponding group
velocity, the guided waves corresponding to the $K_3$ and $K_4$ points propagate in opposite
directions.

Importantly, as shown in \figref{fig:ValleyChirality}(a), the chirality of the upper valley sate at
the $K_1$ point has opposite value as compared to the chirality of the upper valley sate at the
$K_2$ point. Consequently, the valley-Hall topological edge-mode at the $K_3$ and $K_4$ points have
opposite chirality. A consequence of these properties of the chirality and group velocity at the
$K_3$ and $K_4$ points is the unidirectional-propagation property of the topological waveguide
modes of our graphene nanohole plasmonic crystal waveguide.

\subsection{Topological features of the plasmon propagation}\label{sec:TopoPropag}
\begin{figure}[t!]\centering
\includegraphics[width=\columnwidth]{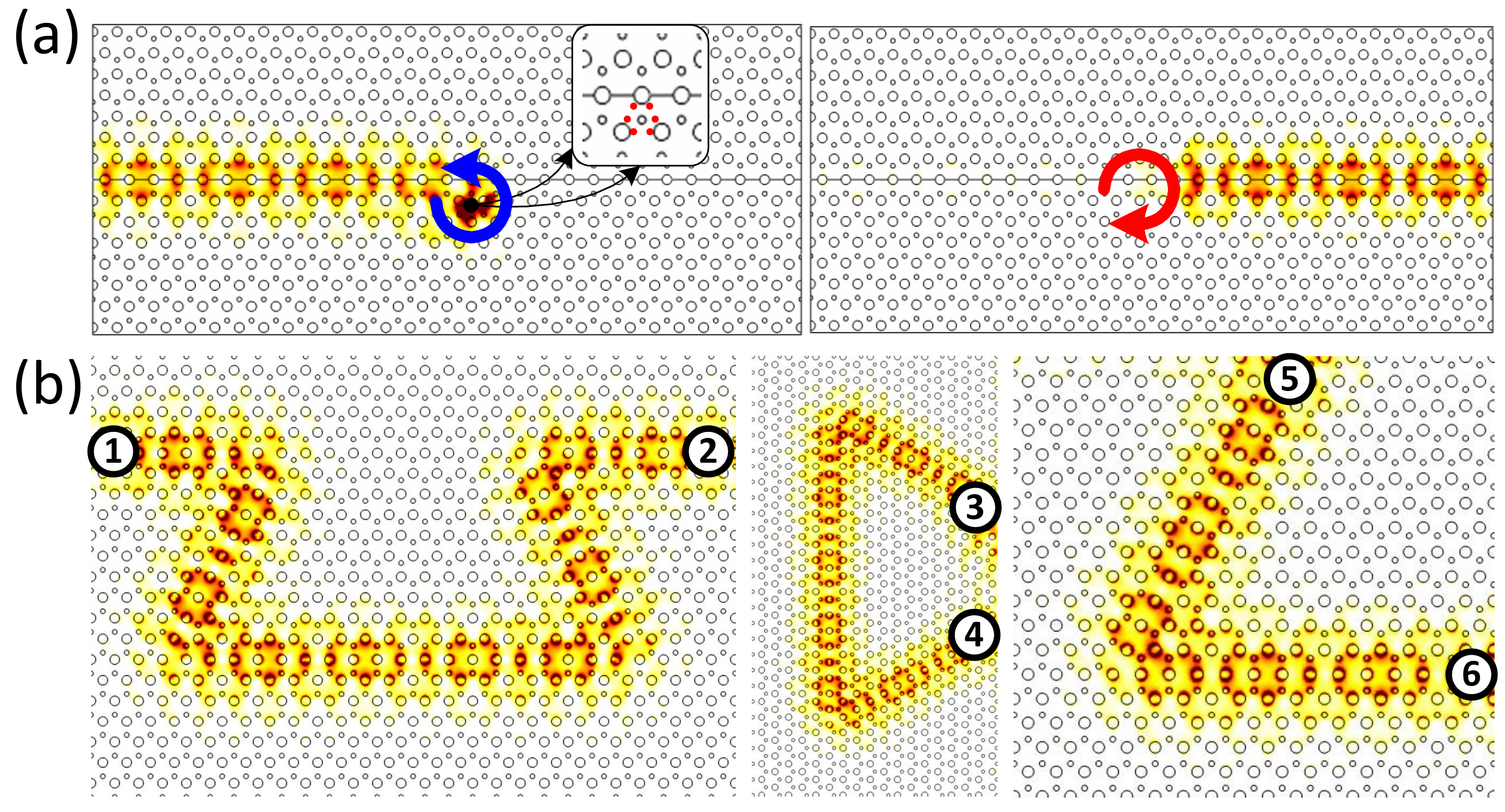}
\caption{Topological features of the topological waveguide mode propagation, including the
unidirectional and the backscattering-immune propagation, in graphene nanohole plasmonic crystal
waveguides. (a) Unidirectional propagation along the $-x$ direction (left panel) and a
unidirectional propagation along the $+x$ direction (right panel) are achieved by placing in the
topological waveguide optical sources that generate RCP light and a LCP light, respectively, at the
frequency of $\nu_0=\SI{11.5}{\tera\hertz}$. (b) Bend-immune propagation in three distinct sharply curved domain-wall interfaces: $U$ bend in the left panel, $C$ bend in the middle panel, and $L$ bend in the right panel. Backscattering-immune propagation is observed, meaning that the proposed graphene nanohole
plasmonic crystal waveguide modes are robust against bend perturbations.}
\label{fig:UniPropagation}
\end{figure}

In order to further investigate the topological features of plasmon propagation in the proposed
graphene nanohole plasmonic crystal waveguide, we considered the propagation properties of several
graphene plasmonic crystal waveguide modes. First, in order to confirm the topologically-protected
unidirectional propagation feature of the waveguide modes, a straight graphene plasmonic crystal
waveguide is studied, as per \figref{fig:UniPropagation}(a). As shown in the left panel of this
figure, a source located at the middle of the graphene plasmonic crystal waveguide generates
right-circularly polarized monochromatic light with frequency of $\nu_0=\SI{11.5}{\tera\hertz}$. In
our study, a source that emits circularly polarized light is implemented by placing at the corners
of a hexagon six electric dipoles, which are marked by red dots in the inset of
\figref{fig:UniPropagation}(a). The phase difference between neighboring electric dipoles is set to
be $\pm\pi/3$ so as to selectively excite a RCP or LCP phase vortex, respectively.

The chirality of the topological waveguide mode at the $K_3$ ($K_4$) point is LCP (RCP).
Consequently, only the topological waveguide mode at the $K_4$ point could be excited in the left
panel of \figref{fig:UniPropagation}(a), if we use as excitation a RCP light source. Moreover,
since the group velocity of the topological waveguide mode at the $K_4$ point is negative in
\figref{fig:BandDiagInterface}(c), the generated optical field will propagate along the --$x$
direction. As a consequence of these two features, a unidirectional propagation of plasmonic waves
is achieved, which is illustrated in the left panel of \figref{fig:UniPropagation}(a). Similarly,
as demonstrated in the right panel of \figref{fig:UniPropagation}(a), unidirectional propagation
along $+x$ direction of plasmonic waves with LCP chirality can be achieved in this graphene
nanohole plasmonic crystal waveguide if one uses as excitation a source that generates
left-circularly polarized light.

In addition to the unidirectional propagation, the topological feature of backscattering-immune propagation
in the proposed graphene nanohole plasmonic crystal waveguide is also validated. As illustrated in
\figref{fig:UniPropagation}(b), three sharply curved domain-wall interfaces with different shapes ($U$ bend in the left panel,
$C$ bend in the middle panel, and $L$ bend in the right panel) are used to test the backscattering-immune propagation in the proposed graphene nanohole plasmonic crystal waveguides. In the
case of the $U$ bend, monochromatic light with frequency $\nu_0$ is launched at the waveguide
port \textcircled{1}, whereas the transmitted power is collected at the waveguide port
\textcircled{2}. The corresponding near-field distribution in the left panel of
\figref{fig:UniPropagation}(b) shows that the light propagation is topologically-protected, as it
is immune to several sharp-bend corners. This feature is particularly promising for the design of
photonic time-delay lines. Similar backscattering-immune propagation can also be observed in the $C$ and
$L$ bends. In the case of the $C$ bend, the feeding waveguide port \textcircled{3} and the
output waveguide port \textcircled{4} are placed at the same side of the crystal. Due to the
topological protection, a backscattering-immune propagation is achieved in this case, too. This remarkable
feature can be used in the design of a photonic device that reverses the direction of light
propagation. In the case of the $L$ bend, the feeding waveguide port \textcircled{5} is placed at
the top side of the crystal, whereas the output waveguide port \textcircled{6} is located at its
right side. Similarly to the phenomena observed in the cases of the $U$ and $C$ bends, in this
case, too, the backscattering of light is efficiently suppressed. This property can be used to the
design of a photonic lateral diverter.
\begin{figure}[t!]\centering
\includegraphics[width=\columnwidth]{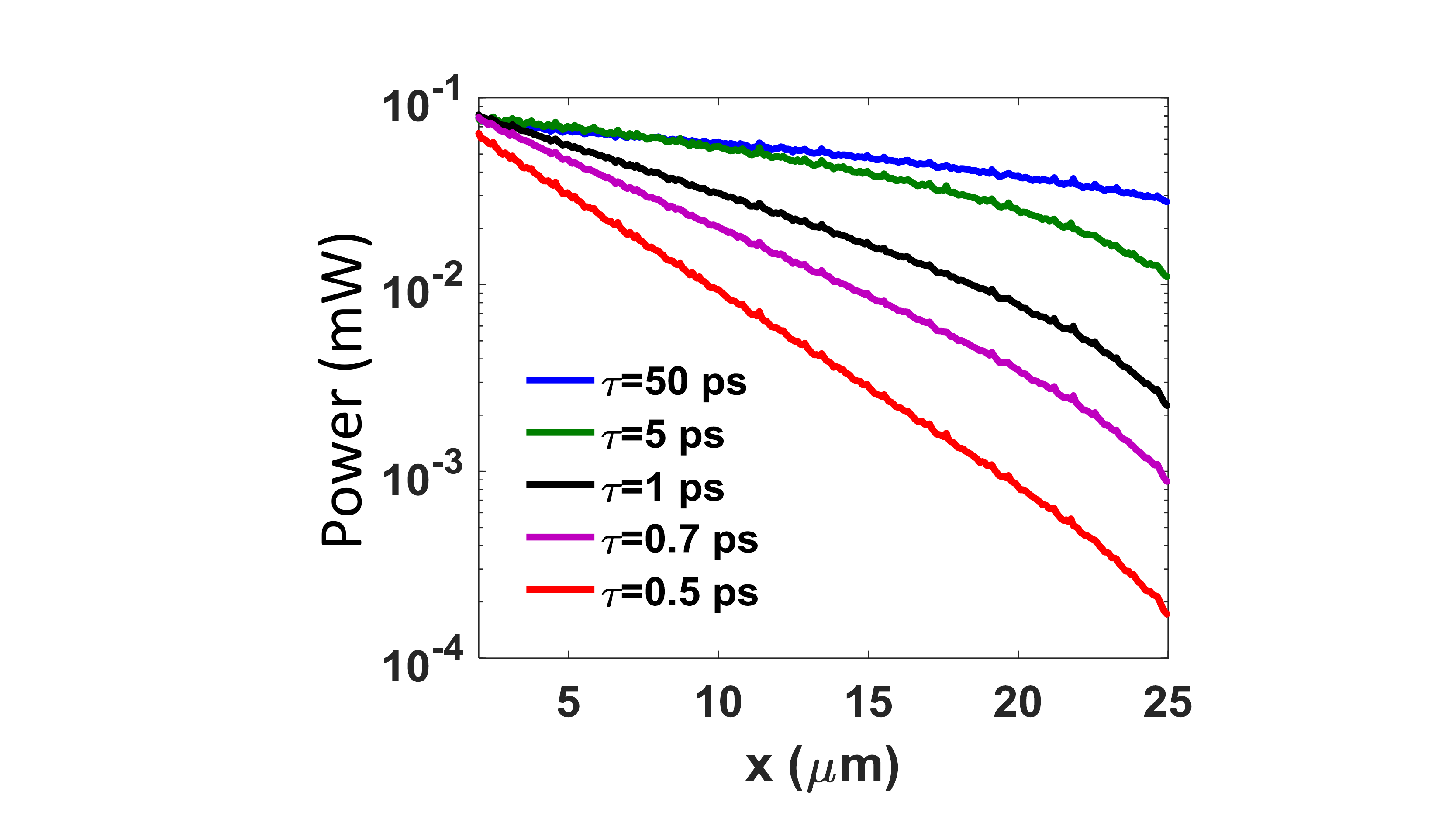}
\caption{Dependence of the modal power on the propagation distance, determined for different values
of the lifetime of the graphene plasmon.} \label{fig:PowerLoss}
\end{figure}

\subsection{The effect of graphene loss on the topological light propagation}\label{sec:LossEffect}
As shown in \figref{fig:BandDiagInfCell}(c), the first bandgap of the graphene nanohole plasmonic
crystals is located at frequencies that are far below the light cone. Thus, the valley-Hall
topological edge modes inside this bandgap are guided modes, as depicted in
\figref{fig:BandDiagInterface}(c), and therefore the radiation losses vanish. Moreover, the
plasmonic field is tightly confined at the domain-wall interface, which can lead to increased
optical losses due to the intrinsic loss of graphene. Typically, the lifetime of graphene plasmons
due to plasmon-phonon scattering varies from \SIrange{0.1}{1}{\pico\second} \cite{la14AcsNano}.
However, recently it has been demonstrated that the lifetime of graphene plasmons can be increased
to about \SI{3}{\pico\second} if the exfoliated graphene is placed onto a boron nitride substrate
\cite{dyml10NatNano}. Moreover, the lifetime of graphene plasmons under a strong static magnetic
field can be as large as \SI{50}{\pico\second}, as demonstrated in \cite{yllz12NanoLett,pwbp13PRL}.
\begin{figure}[t!]\centering
\includegraphics[width=\columnwidth]{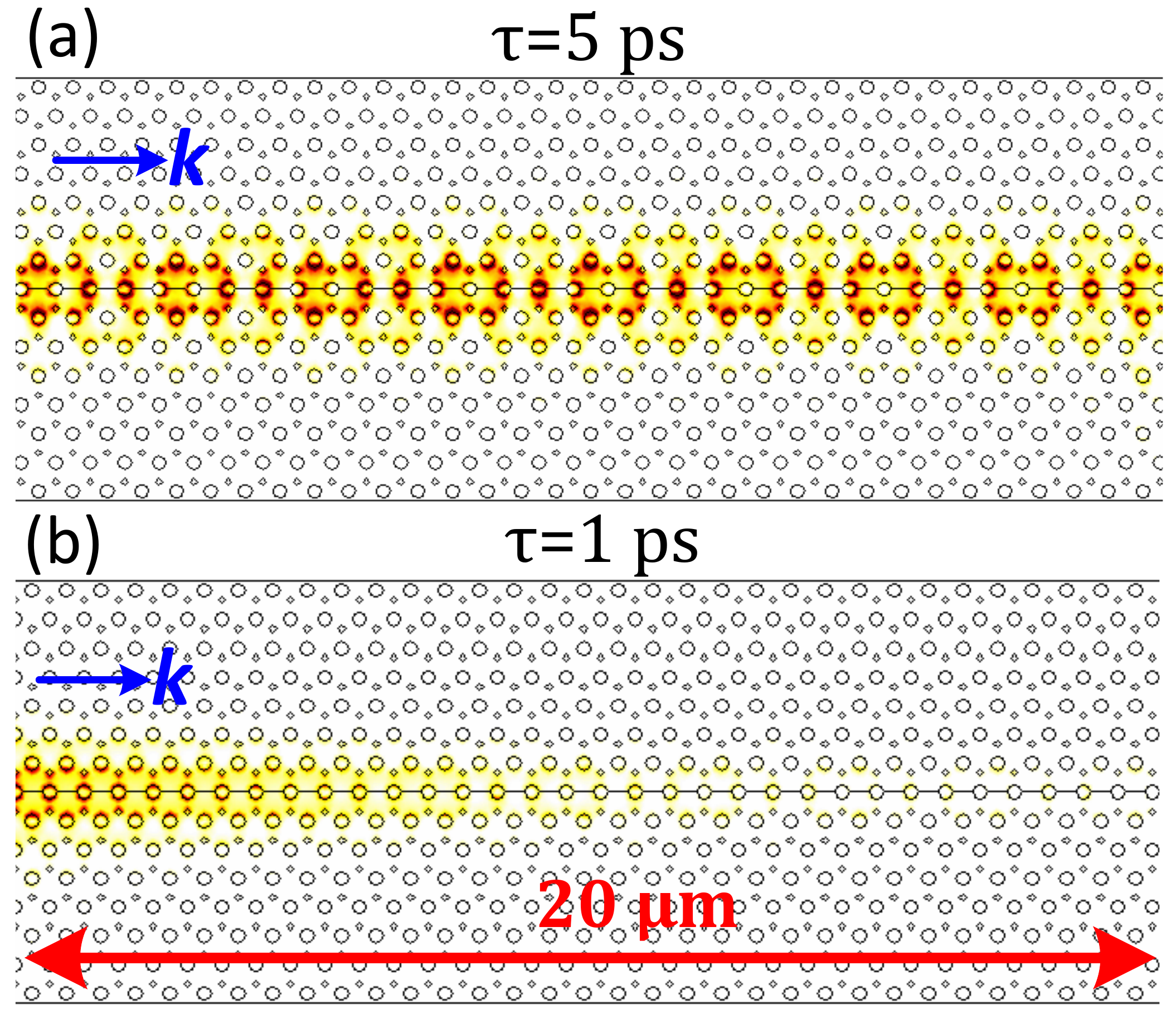}
\caption{(a), (b) Spatial distribution of the electric field intensity, $|E|$, determined for a
lifetime of graphene plasmons of $\tau=\SI{5}{\pico\second}$ and $\tau=\SI{1}{\pico\second}$,
respectively.} \label{fig:FieldLoss}
\end{figure}

Guided by these experimental considerations, we have investigated the effect of the intrinsic loss
of graphene on the characteristics of the valley-Hall topological plasmon propagation in a graphene
nanohole plasmonic crystal waveguide. As shown in \figref{fig:PowerLoss}, the dependence of the
power of the topological waveguide modes on the propagation distance is evaluated via a surface
integral of the Poynting vector over a cross section in the $y-z$ plane. Owing to the intrinsic
loss of graphene, as expected, the optical power of the waveguide mode decreases as the propagation
distance increases. The power decay rate can be quantified by the characteristic loss length,
$L_{loss}$, defined as $L_{loss}=1/\alpha$, where the loss coefficient $\alpha$ is defined by the
relation:
\begin{equation}\label{alpha}
    P(x)=P(0)e^{-\alpha x}.
\end{equation}
For our topological waveguide mode, $L_{loss}=$~\SIlist{27;20;8.3;5.9;4}{\micro\meter} when
$\tau=$~\SIlist{50;5;1;0.7;0.5}{\pico\second}, respectively.

A further illustration of the influence of the intrinsic loss of graphene on the plasmon
propagation in the proposed graphene plasmonic crystal waveguide is provided by the electric field
distributions presented in \figref{fig:FieldLoss}. In both cases the propagation length was equal
to \SI{20}{\micro\meter}, whereas the lifetime of the graphene plasmons was
$\tau=\SI{5}{\pico\second}$ in \figref{fig:FieldLoss}(a) and $\tau=\SI{1}{\pico\second}$ in
\figref{fig:FieldLoss}(b). The results of these numerical simulations clearly illustrate the
dramatic effect that the intrinsic graphene losses has on the characteristic propagation length of
the topological waveguide modes investigated in this work.

\section{Conclusion}\label{sec:Summary}
In summary, we have proposed for the first time a novel valley-Hall topological plasmonic crystal
waveguide formed at the interface between two mirror-symmetric graphene crystals placed in close
contact. After an optimization of the structure of the crystals, a wide topological bandgap
containing an interfacial topological waveguide mode is achieved. Using full-model simulations, the
band diagram of this graphene plasmonic crystal has been investigated, and a topological edge band
was observed inside the bandgap. More importantly, we show that this topological edge band can
exist at extremely deep-subwavelength scale, namely for $\lambda/a>40$. Moreover, we demonstrate
unidirectional and backscattering-immune propagation of the proposed graphene plasmonic crystal waveguide
modes, key features that highlight their topological nature. In addition, the effect of the
intrinsic loss of graphene on the valley-Hall topological propagation in the proposed graphene
nanohole plasmonic crystal waveguide has been investigated, too. We found that characteristic loss
lengths of tens of micrometers can be achieved provided that the plasmon relaxation time is
$\gtrsim\SI{1}{\pico\second}$. These important properties of valley-Hall topological plasmons could
play an important r\^{o}le in the development of highly-integrated and robust photonic crystal
waveguides at deep-subwavelength scale.

\section*{Acknowledgments}
The authors acknowledge the use of the UCL Legion High Performance Computing Facility
(\verb"Legion@UCL"), and associated support services, in the completion of this work.

\ifCLASSOPTIONcaptionsoff
  \newpage
\fi

\end{document}